\documentclass[reprint,aps,prl,showpacs,preprintnumbers,amsmath,amssymb,superscriptaddress]{revtex4-1}

\usepackage{graphicx}
\usepackage{dcolumn}
\usepackage{bm}
\usepackage{wasysym}
\usepackage{xspace}
\usepackage{amssymb}
\usepackage[mediumqspace, thinspace, amssymb]{SIunits}
\usepackage{color}

\newcommand{\dd}{\textrm{d}}

\newcommand{\be}{\begin{equation}}
\newcommand{\ee}{\end{equation}}

\newcommand{\bea}{\begin{eqnarray}}
\newcommand{\eea}{\end{eqnarray}}

\newcommand{\fst}{\quad\textrm{.}}%
\newcommand{\com}{\quad\textrm{,}}%

\newcommand{\2}{(2~1~3-$\tau$)\xspace}

\begin{document}

\title{Itinerant and localized magnetization dynamics in antiferromagnetic Ho}

\author{L. Rettig}
\email[Corresponding author: ]{laurenz.rettig@psi.ch}
\affiliation{Swiss Light Source, Paul Scherrer Institut, 5232 Villigen PSI, Switzerland}
\author{C. Dornes}
\affiliation{Institute for Quantum Electronics, Physics Department, ETH Z\"urich, 8093 Z\"urich, Switzerland}
\author{N. Thielemann-K\"uhn}
\affiliation{Helmholtz-Zentrum Berlin f\"ur Materialien und Energie GmbH, Albert-Einstein-Stra\ss e 15, 12489 Berlin, Germany}
\affiliation{Institut f\"ur Physik und Astronomie, Universit\"at Potsdam, Karl-Liebknecht-Stra\ss e 24/25, 14476 Potsdam-Golm, Germany}
\author{N. Pontius}
\affiliation{Helmholtz-Zentrum Berlin f\"ur Materialien und Energie GmbH, Albert-Einstein-Stra\ss e 15, 12489 Berlin, Germany}
\author{H. Zabel}
\affiliation{Institute for Experimental Physics, Ruhr-Universit\"at Bochum, 44780 Bochum, Germany}
\author{D. L. Schlagel}
\affiliation{Division of Materials Sciences and Engineering, Ames Laboratory, Ames, Iowa 50011, USA}
\author{T. A. Lograsso}
\affiliation{Department of Materials Science and Engineering, Iowa State University, Ames, IA 50011, USA}
\author{M. Chollet}
\affiliation{Linac Coherent Light Source, SLAC National Accelerator Laboratory, 2575 Sand Hill Road, Menlo Park, California 94025, USA}
\author{A. Robert}
\affiliation{Linac Coherent Light Source, SLAC National Accelerator Laboratory, 2575 Sand Hill Road, Menlo Park, California 94025, USA}
\author{M. Sikorski}
\affiliation{Linac Coherent Light Source, SLAC National Accelerator Laboratory, 2575 Sand Hill Road, Menlo Park, California 94025, USA}
\author{S. Song}
\affiliation{Linac Coherent Light Source, SLAC National Accelerator Laboratory, 2575 Sand Hill Road, Menlo Park, California 94025, USA}
\author{J. M. Glownia}
\affiliation{Linac Coherent Light Source, SLAC National Accelerator Laboratory, 2575 Sand Hill Road, Menlo Park, California 94025, USA}
\author{C. Sch\"u{\ss}ler-Langeheine}
\affiliation{Helmholtz-Zentrum Berlin f\"ur Materialien und Energie GmbH, Albert-Einstein-Stra\ss e 15, 12489 Berlin, Germany}
\author{S. L. Johnson}
\affiliation{Institute for Quantum Electronics, Physics Department, ETH Z\"urich, 8093 Z\"urich, Switzerland}
\author{U. Staub}
\affiliation{Swiss Light Source, Paul Scherrer Institut, 5232 Villigen PSI, Switzerland}

\date{\today}

\begin{abstract}

Using femtosecond time-resolved resonant magnetic x-ray diffraction at the Ho $L_3$ absorption edge, we investigate the demagnetization dynamics in antiferromagnetically ordered metallic Ho after femtosecond optical excitation. Tuning the x-ray energy to the electric dipole (E1, $2p\rightarrow 5d$) or quadrupole (E2, $2p \rightarrow 4f$) transition allows us to selectively and independently study the spin dynamics of the itinerant $5d$ and localized $4f$ electronic subsystems via the suppression of the magnetic \2 satellite peak. We find demagnetization timescales very similar to ferromagnetic $4f$ systems, suggesting that the loss of magnetic order occurs via a similar spin-flip process in both cases. The simultaneous demagnetization of both subsystems demonstrates strong intra-atomic $4f$-$5d$ exchange coupling.  In addition, an ultrafast lattice contraction due to the release of magneto-striction leads to a transient shift of the magnetic satellite peak.

\end{abstract}



\maketitle

The manipulation of magnetic order by ultrashort light pulses is of fundamental interest in solid state research and promises high technological relevance. Since the discovery of the demagnetization of Ni in $<\unit{1}{ps}$ almost two decades ago~\cite{Beaurepaire1996}, the ultrafast magnetization dynamics of ferromagnetic systems has been intensely studied both experimentally and theoretically~\cite{Wang2006,Stamm2007,Battiato2010,Koopmans2010,Mathias2012}; for a review see~\cite{Kirilyuk2010,Kirilyuk2013}. In particular the phenomenon of ultrafast magnetization reversal recently observed in ferrimagnetic lanthanide transition metal intermetallics~\cite{Stanciu2007,Radu2011,Ostler2012,Graves2013,Kirilyuk2013,Hassdenteufel2013} has attracted much attention. In the center of this intriguing effect lies the magnetic exchange interaction between localized $f$ moments in the rare-earth atoms, and the itinerant transition metal $d$-electrons. This interaction entails a complex interplay of these two spin reservoirs, such as a transient ferromagnetic state observed during the magnetization reversal in ferrimagnetic FeCoGd~\cite{Radu2011} or momentum transfer between Co and Gd-rich areas~\cite{Graves2013}.

In the rare-earth metals, the magnetic exchange interaction between the large localized moments of their open $4f$ shell is mediated by the indirect Ruderman-Kittel-Kasuya-Yosida (RKKY) interaction via the itinerant $5d6s$ electrons, leading to a parallel alignment of the two subsystems. Depending on the details of the band structure, this interaction leads to a variety of magnetically ordered ground states, ranging from ferromagnetic alignment in Gd and Tb to complex antiferromagnetic (AFM) structures in the heavier rare earths. As optical excitation directly interacts with the valence electrons and not with the localized $4f$ states, these systems  present an ideal case to study the $4f-5d$ interaction directly in the time domain, by separately investigating the dynamics of these two subsystems. While early experiments using x-ray magnetic circular dichroism (XMCD) and magneto-optical Kerr effect (MOKE) on the ferromagnetic lanthanides Gd and Tb found similar demagnetization timescales of $4f$ and $5d$ electrons~\cite{Wietstruk2011}, more recent time-resolved photoemission work found a transient decoupling of the two subsystems in Gd~\cite{Frietsch2015}. However, so far no experiment could directly compare the dynamics of the different spin subsystems using the same observable in a single experiment, and conclusions relied on models and the comparison of different experimental approaches. Furthermore, very little is known about the magnetization dynamics in antiferromagnetic lanthanides, which might provide important insight for the understanding of all-optical magnetization switching in FeCoGd-type ferrimagnets.

In this Letter we investigate the ultrafast demagnetization dynamics of ordered itinerant $5d$ and localized $4f$ moments in antiferromagnetic Ho metal directly, and in a single experiment. Femtosecond time-resolved resonant magnetic x-ray diffraction allows us to separately investigate the dynamics of the $4f$ and $5d$ electrons by choosing either a dipole (E1) or quadrupole (E2) transition in the resonant process. We find a simultaneous demagnetization of $4f$ and $5d$ electrons, demonstrating a strong intra-atomic exchange coupling, while the similarity of the demagnetization dynamics to $4f$ ferromagnets suggests a similar demagnetization process. In addition, an ultrafast shift of the magnetic satellite peak position is attributed to a lattice contraction due to the release of magneto-striction during the demagnetization process.

\begin{figure}[tb]
\includegraphics[width=8.6cm]{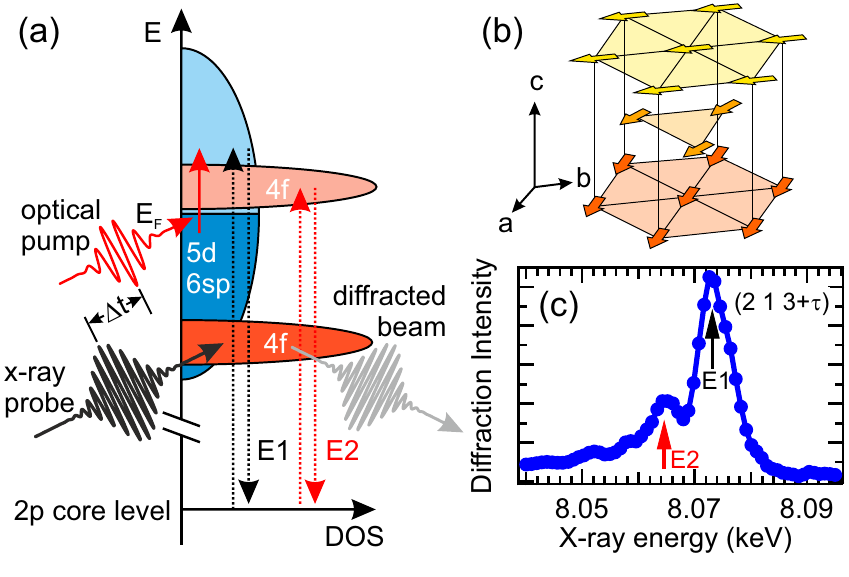}  
\caption{
\label{fig:fig1}
(color online) (a) Schematic energy level diagram and experimental scheme. The resonant x-ray scattering process selectively probes the delocalized $5d$ electrons (E1) and the localized $4f$ electrons (E2) exited by the optical Laser. (b) Crystal structure and magnetic ordering of Ho. The atomic moments (arrows) order ferromagnetically in the $a$/$b$ planes and in an antiferromagnetic spin helix with period $\tau^{-1}$ along the $c$ axis. (c) Resonant x-ray diffraction intensity of the (2~1~3+$\tau$) magnetic satellite peak as a function of incident x-ray energy across the Ho $L_3$ edge. Dipole (E1) and quadrupole (E2) transitions are indicated.
}
\end{figure}

In Ho metal the 3 ($5d6s$) electrons hybridize to form the delocalized, partly occupied valence band structure, whereas the 10 $4f$ electrons remain localized at the atoms and split into occupied and unoccupied manifolds, see Fig.~\ref{fig:fig1}(a). The large experimental magnetic moment of $\approx\unit{11.2}{\mu_B}$ per atom~\cite{JensenMackintosh1991} originates mostly from the large spin and orbital moments of the partially filled $4f$ shell. Below the N\'eel temperature $T_N\approx\unit{133}{K}$, Ho undergoes an antiferromagnetic ordering into a spin helix structure along the $c$ axis with wave vector $\tau\sim 0.3 c^*$ [Fig.~\ref{fig:fig1}(b)]. The length of the spin helix increases with lower temperature and finally locks in at a value of $\tau=1/6$ below $\unit{20}{K}$~\cite{Koehler1966}.

Resonant x-ray diffraction is a direct probe for studying AFM structures, since it is able to directly resolve the atomic-scale pattern of the ordered moments~\cite{Gibbs1988, Schuessler-Langeheine2001, Scagnoli2011, Fink2013}, where the AFM spin helix manifests as magnetic satellite peaks at ($H~K~L\pm\tau$). The resonant x-ray diffraction process at the Ho $L_3$ absorption edge involves virtual transitions between $2p$ core levels and unoccupied valence states, drastically enhancing the sensitivity to the magnetic ordering of the valence states involved in the transition. Thus, by choosing either an electric dipole (E1) or quadrupole (E2) transition, the $5d$ and the $4f$ electrons can be addressed separately due to the respective selection rules~\cite{Gibbs1991, Bouchenoire2009}, as depicted in Fig.~\ref{fig:fig1}(a). As these two transitions are slightly separated in energy, a small modification of the x-ray energy allows us to individually study the magnetization dynamics of the $4f$ and $5d$ electrons independently at the same wave vector.

Time-resolved resonant x-ray diffraction experiments of the magnetic Ho \2 satellite peak were carried out at the X-ray pump-probe (XPP) instrument~\cite{Chollet2015} of the Linac Coherent Light Source (LCLS) free electron laser~\cite{White2015}. The holmium single crystal was excited by $\unit{1.5}{eV}$ laser pulses of $\unit{50}{fs}$ pulse duration at a repetition rate of $\unit{120}{Hz}$, incident at $\unit{5}{^\circ}$ to the surface plane. The energy of the x-ray probe pulses  (pulse duration $\sim\unit{30}{fs}$) was tuned around the Ho $L_3$ edge at an energy of $\unit{8.07}{keV}$ by a thin diamond double crystal monochromator. The diffracted x-rays from each single shot were detected using the Cornell SLAC Pixel Array Detector (CSPAD)~\cite{Hart2012}. The pump-probe arrival time jitter was corrected for shot-by-shot using the spectrally encoding timing tool~\cite{Harmand2013}, yielding a total time-resolution of $\unit{80}{fs}$. A grazing incidence of $\unit{0.5}{^\circ}$ of the x-ray pulses was used to reduce the effective probe depth of the x-rays to match the optical penetration depth of $\lambda_\mathrm{opt}\sim\unit{20}{nm}$~\cite{Krizek1975}. The sample was held at a temperature of $\unit{100}{K} < T_N$ using a cryogenic nitrogen blower throughout the experiments. Static resonant x-ray diffraction experiments characterizing the magnetic order and resonance spectra were performed using a 5-axis surface diffractometer at the X04SA beamline at the Swiss Light Source.

Fig.~\ref{fig:fig1}(c) shows the absorption corrected resonant x-ray diffraction intensity of the magnetic (2~1~3+$\tau$) satellite peak as a function of incident x-ray energy near the Ho $L_3$ edge (which is qualitatively the same for the \2 peak). The spectrum shows two prominent peaks at $\unit{8.064}{keV}$ and $\unit{8.072}{keV}$, below and above the Ho $L_3$ absorption edge at $\unit{8.070}{keV}$, representing a strong resonant enhancement of the magnetic diffraction signal. These two features originate from the electric quadrupole (E2) and electric dipole (E1) transitions in the resonant scattering process, probing the ordered localized $4f$ and itinerant $5d$ moments, respectively~\cite{Bouchenoire2009}. 

\begin{figure}[tb]
\includegraphics[width=8.6cm]{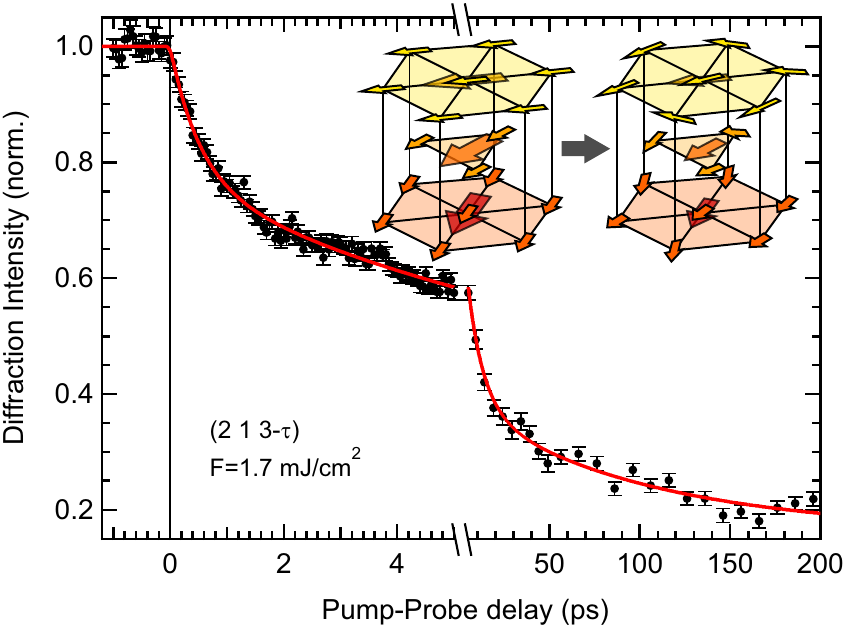}  
\caption{
\label{fig:fig2}
(color online) Time-resolved magnetic x-ray diffraction intensity of the \2 satellite peak at X-ray energy $h\nu=\unit{8.072}{keV}$ as a function of pump-probe delay with an absorbed fluence of $F=\unit{1.7}{mJ/cm^2}$. Error bars are standard errors of the x-ray shot distribution, and the solid line is a three-step fit (see text). Inset: Illustration of the demagnetization process: The average magnetic moment of each layer (large arrows) is reduced by spin-flip scattering.
}
\end{figure}

We first concentrate on the magnetism of the itinerant $5d$ electrons, which are directly excited by the optical pump pulse. 
Its dynamics are probed by the normalized time-dependent diffraction signal $I(t)/I_0$ at the energy of the dipole (E1) transition, where $I_0$ is the intensity before excitation, shown in Fig.~\ref{fig:fig2}. Upon excitation, we observe an initial fast drop of diffraction intensity by $\sim30\%$ within the first picosecond, followed by a further reduction of the intensity on a much slower timescale. After $\unit{200}{ps}$, the diffraction signal is reduced to $\sim 20\%$. In order to extract the different timescales of the demagnetization process, the normalized intensity, which is proportional to the square of the ordered magnetic moments (staggered magnetization), is fit to a phenomenological function consisting of three exponential decays:
\be
I(t)/I_0 = \left[1 - \sum_{i=1}^3\Theta(t) A_i (1-e^{-t/\tau_i}) \right]^2\fst
\label{eqn:fitfunc}
\ee
Here, $A_{1,2,3}$ and $\tau_{1,2,3}$ are the amplitudes and time constants of three demagnetization components, and $\Theta(t)$ is the Heaviside function. A fit to Eq.~\eqref{eqn:fitfunc}, convolved by a Gaussian with a FWHM corresponding to the experimental time resolution of 80 fs is shown in Fig.~\ref{fig:fig2} as a red line and reproduces the data well. The fit yields the demagnetization time-constants $\tau_1 = \unit{0.56\pm0.09}{ps}$, $\tau_2 = \unit{9.5\pm2.2}{ps}$ and $\tau_3 = \unit{119\pm92}{ps}$, and the demagnetization amplitudes $A_1 = 0.12\pm0.01$, $A_2 = 0.25\pm0.04$ and $A_3 = 0.23\pm0.04$.

Such a demagnetization dynamics involving more than one distinct timescale has been previously observed in ferromagnetic rare-earth metals and alloys~\cite{Sultan2011,Wietstruk2011,Sultan2012,Carley2012,Eschenlohr2014,Teichmann2015}. Indeed, the two timescales $\tau_1$ and $\tau_2$ observed here in antiferromagnetic Ho are remarkably close to the demagnetization of ferromagnetic Tb, where a two-step demagnetization with timescales of $\sim\unit{0.7}{ps}$ and $\sim\unit{8}{ps}$ has been reported~\cite{Wietstruk2011}. These two timescales of the demagnetization have been interpreted in terms of hot-electron-mediated spin-flip scattering and slower phonon-assisted spin-lattice relaxation, respectively~\cite{Koopmans2010, Roth2012}. The further demagnetization with time constant $\tau_3$ is most likely due to heat transport within the probed volume.

\begin{figure}[tb]
\includegraphics[width=8.6cm]{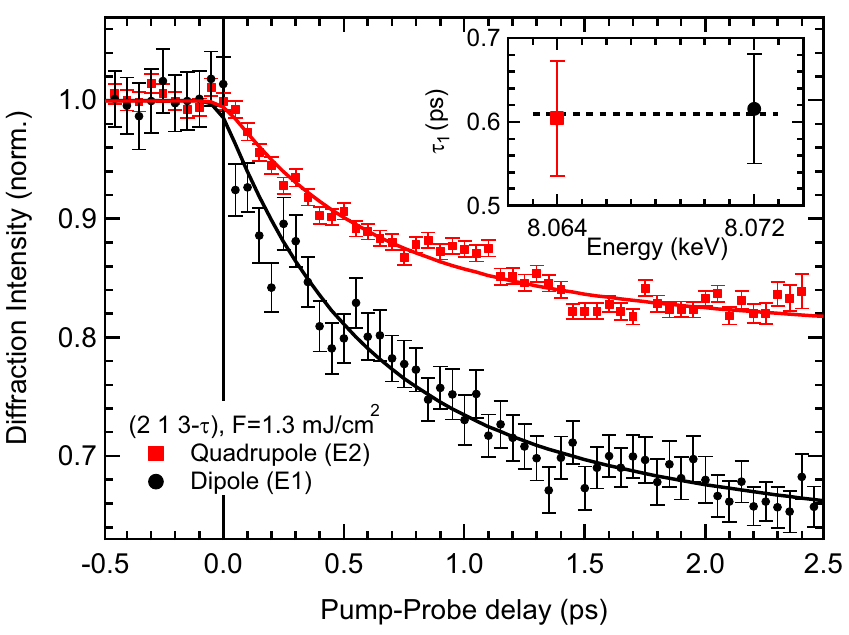}  
\caption{
\label{fig:fig3}
(color online) Time-resolved magnetic x-ray diffraction intensity for $h\nu=\unit{8.064}{keV}$ (quadrupole, red squares) and $h\nu=\unit{8.072}{keV}$ (dipole, black circles) as a function of pump-probe delay at an absorbed pump fluence of $F=\unit{1.3}{mJ/cm^2}$. The inset shows the fast relaxation time constants $\tau_1$ obtained from a fit to the data (solid lines, see text). Error bars are 95\% confidence intervals of the fits.
}
\end{figure}

Further information about the role of the $4f$ and $5d$ electrons and their coupling in the demagnetization process can be gained by  tuning the energy of the resonant x-ray probe pulses across the Ho $L_3$ edge, thereby selectively probing the respective electron systems. Figure~\ref{fig:fig3} shows $I(t)/I_0$ at the energy of the dipole (E1, black) and quadrupole (E2, red) transitions, which independently probe the magnetic ordering of the itinerant $5d$ and of the localized $4f$ spin systems, respectively. Both curves show a very similar fast demagnetization, well described by the  demagnetization behavior observed in Fig~\ref{fig:fig2}. In order to correctly describe the demagnetization amplitudes, the change of the x-ray penetration depth across the Ho $L_3$ edge from $\lambda_\mathrm{x-ray}(E_1)=\unit{15}{nm}$ to $\lambda_\mathrm{x-ray}(E_2)=\unit{40}{nm}$ has to be taken into account. For coherence of the diffracted beam~\cite{Beaud2014} the diffracted intensity can  be written as
\be
I(t)\propto\left|\int_0^\infty M(z,t) \cdot e^{-z/2\lambda_\mathrm{x-ray}}  \dd z\right|^2\com
\label{eqn:2}
\ee
where $M(z,t)$ is the normalized time- and depth-dependent magnetic structure factor (proportional to the staggered magnetization), and its reduction is assumed to depend linearly on the pump energy density at depth $z$, with the time-dependence of Eq.~\eqref{eqn:fitfunc}:
\be
M(z, t) =  1 - \left(\sum_{i=1}^3\Theta(t) A_i (1-e^{-t/\tau_i})\right) \cdot e^{-z/\lambda_\mathrm{opt}} .
\label{eqn:3}
\ee
Fits of Eq. \eqref{eqn:2} and \eqref{eqn:3} to the data shown in Fig.~\ref{fig:fig3} yield maximal fast demagnetization amplitude $A_1^\mathrm{E1}=0.22\pm0.01$ and $A_1^\mathrm{E2}=0.21\pm0.01$, and time constants $\tau_1^\mathrm{E1}=\unit{0.60\pm0.07}{ps}$ and $\tau_1^\mathrm{E2}=\unit{0.62\pm0.08}{ps}$, identical for the two magnetic subsystems within our accuracy. 

The observation of identical demagnetization timescales of $5d$ and $4f$ electrons is intriguing. Whereas the optical excitation directly affects only the small moments of the itinerant conduction electrons ($\mu_{5d}\approx\unit{0.6}{\mu_B}$), the localized $4f$ moments, which carry most of the ordered magnetic moments ($\mu_{4f}\approx\unit{10.6}{\mu_B}$), are only indirectly affected by the pump pulse though intra-atomic $5d$-$4f$ exchange coupling. Therefore, depending on the strength of this coupling, one could expect a faster demagnetization of the $5d$ states. Such a behavior of different demagnetization timescales has been observed e.g. in the demagnetization of the different elements in $3d$/$4f$ alloys~\cite{Radu2011, Graves2013}. The identical demagnetization timescales for the $5d$ and $4f$ electrons in Ho demonstrate a very strong intra-atomic exchange coupling between the two spin systems. This strong coupling efficiently ties the $5d$ moments to the large $4f$ moments and prevents a selective demagnetization of the conduction electrons, leading to the simultaneous demagnetization of both spin systems. Indeed, calculations of the intra-atomic $f$-$d$ exchange coupling constant yield $J_{fd}\sim\unit{70}{meV}$ for Ho~\cite{Ahuja1994}, corresponding to a characteristic time scale of $\sim\unit{10}{fs}$, well within our experimental error bars. Such a strong intra-atomic exchange coupling of itinerant and localized magnetic moments was also discussed for ferromagnetic Gd and Tb~\cite{Wietstruk2011}, suggesting a general behavior in the rare-earth systems. We note, however, that our finding of identical demagnetization timescales of $4f$ and $5d$ electrons are in contrast to the decoupled ultrafast magnetic dynamics recently observed for \emph{occupied} $d$ and $f$ states by time-resolved photoemission in Gd~\cite{Frietsch2015}.

We now turn back to the timescales of the demagnetization. In ferromagnetic systems, the demagnetization rate is generally considered to be limited by the dissipation of angular momentum from the polarized spin system via angular momentum transfer to the lattice~\cite{Koopmans2010, Roth2012} or through spin transport channels~\cite{Battiato2010, Turgut2013, Eschenlohr2013}. In contrast, in an antiferromagnet, the total sublattice magnetizations compensate each other and no net angular momentum needs to be conserved during ultrafast demagnetization. Therefore, demagnetization of AFM systems could potentially be significantly faster than in ferromagnetic systems. Indeed, demagnetization in various strongly correlated antiferromagnetic systems such as Fe pnictides~\cite{Rettig2012, Kim2012}, cupric oxide~\cite{Johnson2012} or Nickelates~\cite{Caviglia2013} have been reported to progress on much faster timescales. 

The similarity of the demagnetization timescales observed here in AFM Ho compared to ferromagnetic lanthanides, and the lack of a significantly faster demagnetization suggests that the demagnetization in Ho occurs via similar processes involving angular momentum dissipation, despite the absent net magnetization. A possible route could be a loss of AFM order by demagnetization of the individual ferromagnetic sublattices along the $a/b$-planes, as sketched in the inset of Fig.~\ref{fig:fig2}, whereas the AFM helical alignment of ferromagnetic planes along the $c$ axis stays constant. Such a scenario suggests that the spin flip scattering mechanisms leading to demagnetization may be shorter in range than the helix period. It may, however, also play a role that the present experiment was carried out near the ordering temperature and that critical slowing down~\cite{Manchon2012, Kimling2014} obscures an otherwise faster dynamics. To clarify this issue, further complementary experiments e.g. at the Ho $M_5$ edge could provide additional insight.

\begin{figure}[tb]
\includegraphics[width=8.6cm]{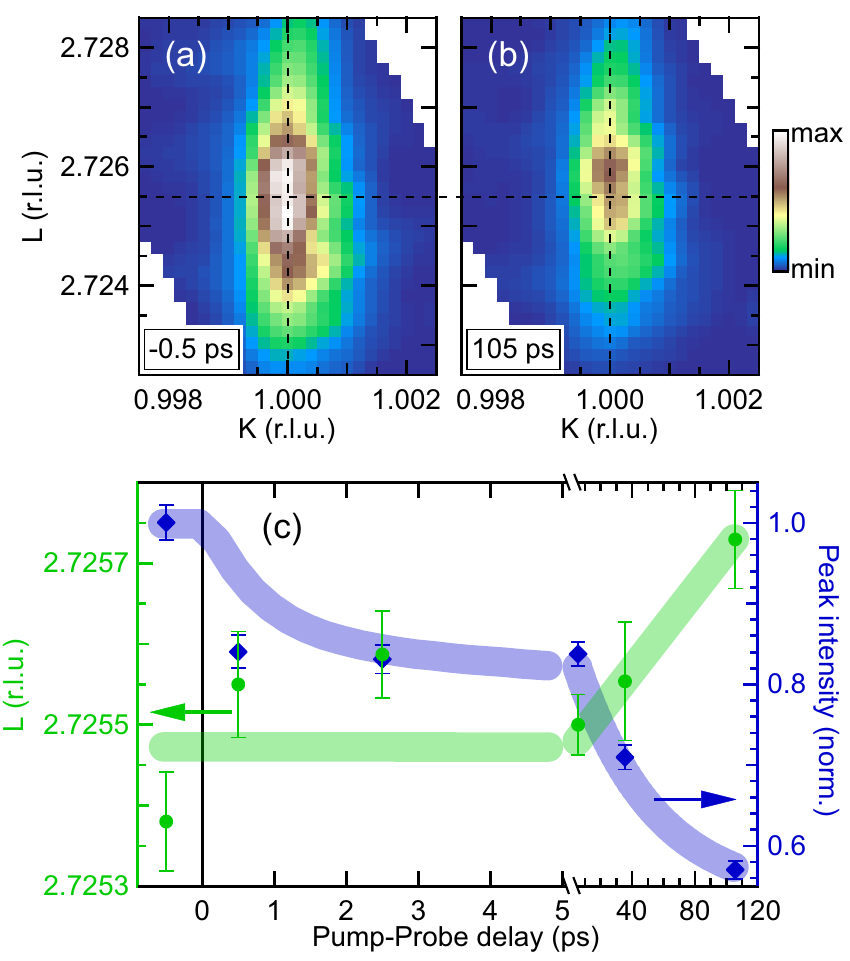}  
\caption{
\label{fig:fig4}
(color online) Reciprocal space maps of the magnetic x-ray diffraction intensity along the $(K,L)$ plane at $H=2$ (a) before and (b) at $\unit{105}{ps}$ after excitation. Dashed lines mark the peak position before excitation. Note the shift towards larger $L$ after excitation. (c) Peak position (circles, left axis) and maximum peak intensity (diamonds, right axis) along the $L$ direction as function of pump-probe delay determined by Gaussian fits to line profiles along the L-direction. Shaded areas are guides to the eyes, and error bars are 95\% confidence intervals of the fits.
}
\end{figure}

Finally, the time-resolved resonant x-ray diffraction also allows us to investigate the dynamics of the transient magnetic structure during demagnetization in reciprocal space. Fig~\ref{fig:fig4}(a) and (b) show a cut of the magnetic diffraction intensity of the \2 satellite in the $(K,L)$ plane, at $H=2$, before, and $\unit{105}{ps}$ after excitation. Apart from the reduction of the diffraction intensity due to the demagnetization, a clear shift of the peak center towards larger $L$ is observed. The time dependence of peak position (green) and intensity (blue) is determined by Lorentzian squared line fits along the $L$ direction, shown for various pump probe delays in Fig.~\ref{fig:fig4}(c), while no change in the peak width (correlation length) is observed.

Such a transient shift of a magnetic satellite peak can in principle have two origins: (i) a change of the ordering vector $\tau$ shifting the satellites relative to the structural peak, or (ii) a change of the crystal lattice constant $c$, shifting the structural peak position along with the satellites. For the first case, the observation of a shift towards \emph{larger} $L$ of the \2 satellite corresponds to a \emph{decrease} of $\tau$ upon excitation. Such a behavior seems unlikely, as it is in contrast to a thermal behavior, where an increase of $\tau$ with increasing temperature is observed~\cite{Koehler1966}. In contrast, an increase in $L$ can be explained by a contraction of the lattice constant $c$ upon excitation, due to the release of magneto-striction, which statically leads to an anomalous expansion of $c$ when entering the magnetic helical phase~\cite{Darnell1963}. 

In conclusion, we investigated the ultrafast demagnetization dynamics in antiferromagnetic Holmium using time-resolved resonant x-ray diffraction at the Ho $L_3$ edge. The demagnetization of the $5d$ electrons proceeds via a three-step demagnetization process on timescales very similar to ferromagnetic $4f$ materials, indicating a similar spin-flip scattering mechanism for the loss of magnetic order in these systems. The demagnetization of $4f$ and $5d$ electrons follows the same time dependence, demonstrating a strong intra-atomic exchange coupling between the two spin systems. The suppression of antiferromagnetic order leads to the release of magneto-striction, which manifests in an ultrafast lattice contraction upon excitation.

\begin{acknowledgments}
This research was carried out on the XPP Instrument at at the Linac Coherent Light Source (LCLS) at the SLAC National Accelerator Laboratory. LCLS is an Office of Science User Facility operated for the US Department of Energy Office of Science by Stanford University. Use of the Linac Coherent Light Source (LCLS), SLAC National Accelerator Laboratory, is supported by the U.S. Department of Energy, Office of Science, Office of Basic Energy Sciences under Contract No. DE-AC02-76SF00515. Static resonant x-ray diffraction experiments were performed at the X04SA Material Science beamline at the Swiss Light Source, Paul Scherrer Institut, Villigen, Switzerland. The Ho single crystal was prepared by the Materials Preparation Center at the Ames Laboratory. The Ames Laboratory is operated for the U.S. Department of Energy by Iowa State University under Contract No. DE-AC02-07CH11358. This work was supported by the NCCR Molecular Ultrafast Science and Technology (NCCR MUST), a research instrument of the Swiss National Science Foundation (SNSF), and the Helmholtz Virtual Institute Dynamic Pathways in Multidimensional Landscapes. HZ acknowledges support through BMBF 05K13PC1. 

\end{acknowledgments}

%

\end{document}